\documentclass[fleqn,twoside]{article}
\usepackage{espcrc2}

\usepackage{graphicx}
\newcommand{\AmS}{{\protect\the\textfont2
  A\kern-.1667em\lower.5ex\hbox{M}\kern-.125emS}}
\newcommand{\beq}{\begin{equation}}
\newcommand{\eeq}{\end{equation}}
\newcommand{\beqa}{\begin{eqnarray}}
\newcommand{\eeqa}{\end{eqnarray}}

\title{Dynamics of ferromagnetic nanomagnets with vortex or single-domain configuration}

\author{A.\ Killinger\address{Inst.\ of Physics II, University of Regensburg, D-93040 Regensburg,
Germany}\thanks{Tel.: +49-941/9432605; Fax: +49-941/9434544; {\it
E-mail address:} andreas.killinger@physik.uni-regensburg.de}, R.\
H{\"o}llinger\addressmark ,\ U.\ Krey\addressmark}

\begin{document}

\begin{abstract} We study the dynamics of flat circular permalloy nanomagnets
for 1.) {\it magnetic vortex} and 2.) {\it single-domain configurations} using
micromagnetic simulation. Dynamical studies for isolated {\it vortex}
structures show that both the vorticity and the polarity of the
out-of-plane component can be switched fast (50-100 ps).
Micromagnetic simulations of the switching process in thin
cylindrical Permalloy(Py) nanoparticles with an initial stable
{\it single-domain} state show nearly homogeneous single-domain behavior
followed by  excitation of spin waves. \vspace{1pc}

\end{abstract}

\maketitle

The magnetism of small ferromagnetic structures has become
increasingly important e.g. for nonvolatile random access memories
(MRAMs) \cite{cow}. Nanostructured Permalloy is a possible
candidate for such devices. As it is an essential feature in the
technology of magnetic recording to read and write magnetic states
as fast as possible we study the dynamics of flat circular
permalloy nanomagnets with magnetic vortex or single-domain
configuration using micromagnetic simulations.

There are four equivalent vortex states since the vorticity
(clockwise/counterclockwise) and the central polarization (up or
down) are independent. The possibility of storing and switching
two bits of information instead of only one makes these vortex
structures quite interesting.

It is known  from \cite{hoell} that the vorticity of a magnetic dot can be
switched in $\approx$ 40 ps with strong enough and short enough
perpendicular out-of-plane field pulses.

In the present work, we consider 1.) the question, 
 whether also the
polarization can be switched reproducibly and as fast as the
vorticity.

Therefore the results of a micromagnetic simulation performed with
OOMMF, \cite{oommf}, on a cylindrical Py dot with diameter 300 nm and
10 nm thickness are shown. We have used  Py parameters
($M_{s}=800000\,\frac{\rm A}{\rm m}$ for the saturation magnetisation,
$A=1.3\ast10^{-11}\,\frac{\rm J}{\rm m}$ for the exchange stiffness and
$\alpha=0.008$ for the Gilbert damping). The initial magnetisation
configuration is a (relaxed) vortex state. We apply a typical out-of-plane field pulse
(100 ps duration) of
$B_{z}=0.2\,{\rm T}$, plus $B_{x}=0.002\,{\rm T}$ to induce torque at
$x=y\approx 0$.
\vspace{-0.5 truecm}
\begin{figure}[h]
\begin{center}
\includegraphics[height=6cm,width=6cm]{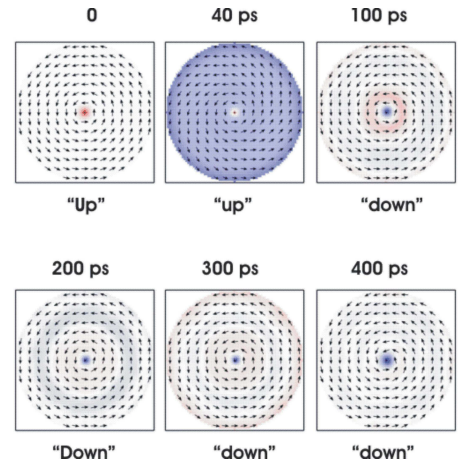}
\caption{\it {\rm (Color):} {Bistable switching of the central polarisation of a Py dot with vortex
structure}}\label{sequ}\vskip 0.2 truecm
\end{center}
\end{figure}
\vspace{-1.5cm}
Fig.\ 1 shows the
magnetic configurations of this dot at different times. After 100 ps
the central polarization flips from "up" to "down".

Thus it is possible to switch the central polarization of a Py dot with
vortex configuration in 100 ps by applying strong {\it out-of-plane} perpendicular
field pulses.  So these Py dots are  candidates for
independent
switching of both 
the chirality, and the central polarisation of {\it vortex
configurations}.-- 2.) In the following the switching dynamics of 
{\it single-domain-states} in a flat circular Py cylinder are considered
(diameter 150 nm, thickness 2 nm, same material as above, with an
additional small in-plane uniaxial anisotropy
$K_u = 500\,{\rm J/m^3}$ with easy axis along the magnetisation
at $t=0$; the initial state is a stable single-domain-state). In the
simulation now an {\it in-plane} field-pulse of $0.05\,{\rm T}$ (strong enough for
precessional switching) perpendicular to
the magnetisation was applied instantaneously for $t\ge 0$ and switched off
at $t \approx 0.1\,{\rm ns}$, just after the reversal of the  magnetisation.
\vspace{-0.6cm}
\begin{figure}[h]
\begin{center}
\includegraphics[height=6cm,width=8cm]{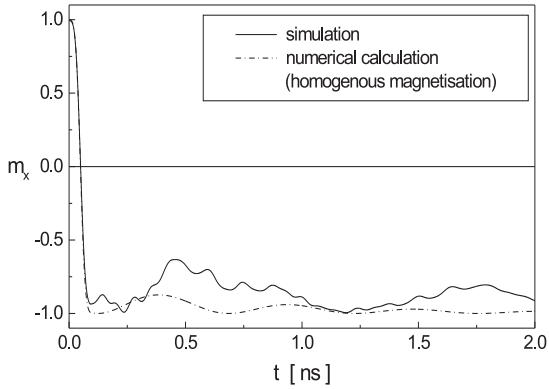}
\vspace{-1.1cm}
\caption{\it {Dynamics of the x-component of the magnetisation}}\label{fig2}
\end{center}
\end{figure}
\vspace{-0.9cm}
Fig.\ 2 shows the time-dependence of the x-com\-ponent of the magnetisation,
compared to a numerical calculation of the same process with fictitiously
$homogenous$ magnetisation (no exchange interactions). 

The main results are:
\begin{itemize}
\item [a)] During the switching process ($t \leq 0.1\,{\rm ns}$)
single-domain-behaviour predominates.
\item [b)] After the reversal dipole-exchange-spinwaves are excited.
\item [c)] The 
 energy of the spin-system decreases exponentially $
\sim \exp(-t/\tau)$ with a time constant $\tau \approx 0.65\,{\rm ns}$.
\end{itemize}
\vspace{-0.9cm}
\begin{figure}[h]
\begin{center}
\includegraphics[height=6cm,width=8cm]{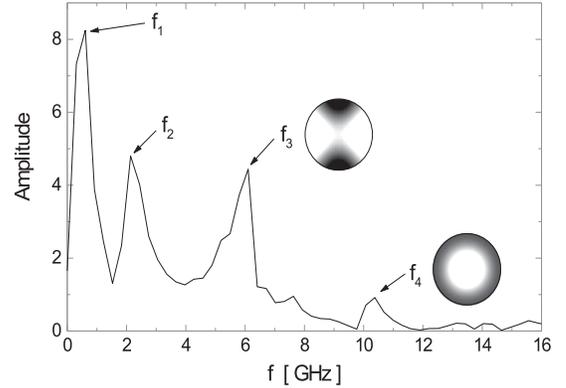}
\vspace{-1.0cm}
\caption{\it {Frequency-spectrum of the z-component of the magnetisation}}
\label{fig3}
\end{center}
\end{figure}
\vspace{-1.0cm}
A Fourier transformation of $M_z(t)$
leads to a 'discrete' frequency-spectrum (fig.\ \ref{fig3}) because of the
edge-conditions (free spins at the edge). The lowest frequencies $f_1$ and
$f_2$ can be explained by uniform precession of the spins in the anisotropy
fields of the above-mentioned uniaxial anisotropy $K_u$ and a (fictitious)
fourfold
anisotropy, generated by the discretization of the simulation volume; 
 the frequencies
$f_3$ and $f_4$ are eigenmodes of the spin-wave-spectrum, the
corresponding eigenfunctions are shown as inset in figure \ref{fig3} (the
greyscale corresponds to the z-component of the magnetisation). A 
theory for these quantized modes is given in \cite{gusl}.



\begin{thebibliography}{99} 
\bibitem{cow} R.\ P.\ Cowburn, D.\ K.\ Koltsov,
A.\ O.\ Adeyeye {\it et al.},\ Phys. Rev. Lett. {\bf 83} (1999) 1042

\bibitem{gusl} K.\ Yu.\ Guslienko, A.\ N.\ Slavin, \ J.\ Appl.\ Phys.\ 87
(2000) 6337
\bibitem{hoell} R.\ H{\"o}llinger, A.\ Killinger, U.\ Krey, \ J. Magn. Magn. Mater. {\bf
261} (2003) 178
\bibitem{oommf} {\it Object Oriented Micromagnetic Framework}, http://math.nist.gov/oommf


\end{thebibliography}
\end{document}